\documentclass{aa}
\usepackage[dvips]{epsfig}

\begin{document}

\title{Spectropolarimetry of the borderline Seyfert 1 galaxy\\ 
       ESO 323-G077
       \thanks{Based on observations obtained at the ESO VLT UT1
(Antu) at Cerro Paranal, Chile (ESO program 66.B-0063).}}
\subtitle{}
\author{H.M.~Schmid\inst{1}
        \and I.~Appenzeller\inst{2}
        \and U.~Burch\inst{1}}

\offprints{H.M.~Schmid, \email{schmid@astro.phys.ethz.ch}}

\institute{Institut f\"ur Astronomie, ETH Zentrum, CH-8092 Z\"urich, 
            Switzerland
\and  Landessternwarte Heidelberg-K\"onigstuhl, 
            D-69117 Heidelberg, Germany }

\date{received; Accepted}

\abstract{We report the detection of high linear polarization
in the bright Seyfert~1 galaxy \object{ESO 323-G077}.
Based on optical spectropolarimetry with FORS1 
at the VLT we find a continuum polarization which ranges from 
2.2~\% at 8300~\AA\ to 7.5~\% at 3600~\AA. Similar amounts 
of linear polarization are found for the broad emission lines, 
while the narrow lines are not polarized.
The position angle of the polarization ($\theta=84^\circ$) 
is independent of the wavelength and found to be perpendicular 
to the orientation of the extended [O\,{\sc iii}] 
emission cone of this galaxy. Within the standard model of
Seyfert nuclei the  observations can be well understood
assuming that this AGN is observed at an inclination angle
where the nucleus is partially obscured and seen mainly indirectly 
in the light scattered by dust clouds within or above the torus and 
the illuminated inner edge of the dust torus itself.
Hence we conclude that ESO 323-G077 is a borderline Seyfert~1 
galaxy which can provide important information on the geometric
properties of active nuclei.
\keywords{galaxies: active -- galaxies: Seyfert --
          galaxies: individual: ESO 323-G077 -- polarization -- scattering }
}
\authorrunning{H.M. Schmid et al.}
\titlerunning{Spectropolarimetry of the Sy~1 galaxy ESO 323-G077}

\maketitle

\section{Introduction}

With $m_{\rm V}=13.2$~mag  
ESO 323-G077 is one of the brightest Seyfert 1
galaxies known. The host galaxy is an SB0 spiral at $z=0.0150$ with a weak 
bar and a triaxal bulge (Greusard et al.~\cite{greusard00}). 
The presence of an active nucleus 
was reported first by Fairall (\cite{fairall86}). This relatively
late discovery may have been due to the galaxy's 
location in the less well surveyed southern sky 
and its relatively low galactic latitude ($b=+22^\circ$). 
Therefore  
ESO 323-G077 has not been included in any of the polarimetric 
surveys of active galaxies like those of e.g. Martin et al.~(\cite{martin83})
or Brindle et al.~(\cite{brindle90a}). No polarimetric
observations have been reported in the literature so far. On the 
other hand, the nuclear colors and the spectrum of ESO 323-G077
indicate substantial reddening (Winkler et al.~\cite{winkler92a}; 
Winkler \cite{winkler92b}), a property often observed in 
high-polarization Seyfert 1 galaxies, such as  
\object{Fairall 51} (Schmid et al.~\cite{schmid01}) or \object{Mrk 231}
(Smith et al.~\cite{smith95}). This prompted us
to obtain for ESO 323-G077 spectropolarimetric observations, which
indeed revealed a high linear polarization.

Scattering polarization is well known to occur in 
Seyfert 2 galaxies and is particularly important for the investigation
of aspect-dependent properties of these objects.
Spectropolarimetric observations of Seyfert 2s often show  
the presence of Seyfert 1 type features in polarized light
and thus provide the most direct evidence for the standard unified
model of AGN (Antonucci \& Miller \cite{antonucci85}; 
Antonucci \cite{antonucci93}).

As pointed out below the high polarization detected for the
ESO 323-G077 appears to be of similar origin as in Seyfert 2 galaxies.
However the inclination angle seems to fall in between the typical
values for Seyfert~1 and Seyfert~2 galaxies. Therefore, this galaxy 
offers the opportunity to investigate the geometric properties of
the obscuring and light collimating torus assumed in the standard
AGN models. Compared to other borderline Seyfert 1 galaxies 
ESO 323-G077 has the additional advantage that the geometric
orientation of the torus can be inferred from the [O\,{\sc iii}]
emission cone which is known to exist in this object from earlier
observations.

Following this introduction we outline in 
Sect.~2 the observations and the data reduction, 
followed by a description of the spectroscopic and polarimetric 
properties of the nucleus in ESO 323-G077 in Sect.~3. In Sect.~4 an
interpretation of the observations is given. 

\begin{table} [t]
\begin{center}       
\caption{ESO 323-G077 observing log. $t_{\rm exp}$ is the total exposure time
of an entire polarization observation comprising four exposures each.}
\label{tab.log}
\begin{tabular}{lllrll} 
\noalign{\hrule\smallskip}
grism & filter & $\lambda$-range & Res. & date 
                                              & $t_{\rm exp}$ \\
      &        & [\AA]          & $\lambda/\Delta\lambda$ & 2001 & [min]  \\   
\noalign{\smallskip\hrule\smallskip}
\noalign{\smallskip \raggedright ESO 323-G077 \smallskip}
G300V & GG 375 & 3770--8600     & 430   & Jan.~29  & 20  \\
\noalign{\smallskip} 
G600B &        & 3480--5850     & 815  & Jan.~29 & 80     \\
\noalign{\smallskip}
G600R & GG 435 & 5250--7410     & 1230  & Jan.~31 & 40     \\
\noalign{\smallskip\noindent \raggedright GSC 07777-00007$^*$ \smallskip}
G300V & GG 375 & 3770--8600     & 430 & Jan.~29 & $\phantom{0} 4$ \\
\noalign{\smallskip\hrule}
\end{tabular}
\end{center}
\noindent
$*$: star, $m_{\rm V}=11.4$~mag, located 30\arcsec\ west 
and 6\arcsec\ north of the ESO 323-G077 nucleus
\end{table}
%

\section{Observations}

Optical spectropolarimetry of ESO 323-G077 was obtained during 
the two nights of January 29 and 31, 2001 with FORS1 
(Appenzeller et al.~\cite{appenzeller98}) at the ESO VLT Unit 
Telescope UT~1  (Antu). Data were taken using the three grisms 
G300V, G600B and G600R as summarized in Table~\ref{tab.log}.
In all cases a slit width of  1\arcsec\ , a slit height of 20\arcsec ,
and a north-south slit orientation was used.
The linear polarization was measured using cycles of
4 exposures with different half-wave plate position angles,
as described in Schmid et al.~(\cite{schmid01}). 
The spectrum was extracted from the central 
10$''$. DIM seeing was always about 1\arcsec.

To determine the Galactic interstellar polarization 
in the direction to ESO 323-G077 we made a short spectropolarimetric 
observation of the nearby star \object{GSC 07777-00007} 

The instrument response curve was determined using wide-slit observations
of the spectrophotometric standard star EG~21. This response curve
was used to carry out an approximate flux calibration of the relative flux
observed through the 1 arcsec slit centered on the nucleus.
However, since the observing conditions were not photometric
these flux values are of uncertain absolute accuracy, although the
relative flux should be correct.
Since we found that the calibrated spectra
obtained with two different grisms taken on Jan.~29 agree rather well
(deviations $<$5~\%), we regard the fluxes obtained for that
night more reliable. The flux obtained from the G600R spectrum taken on 
Jan.~31 is 45~\% lower. Hence we adjusted the
flux of the G600R spectrum to the G300V/G600B scale. 

\begin{figure*}
\begin{center}
\hskip-1cm\hbox{\psfig{figure=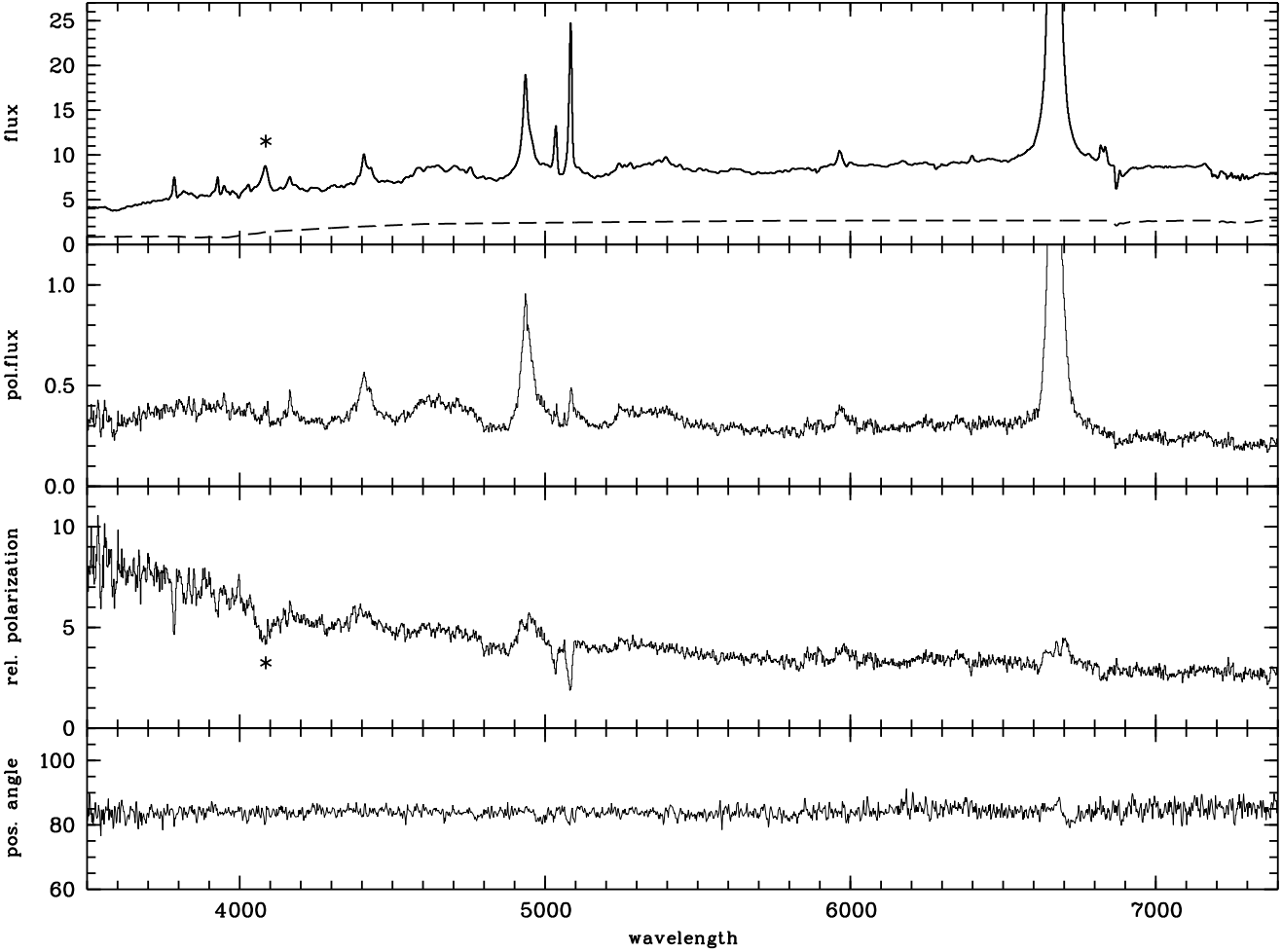,width=18cm}}
\end{center}
\vskip-0.5cm
\caption[] 
{\label{fig.pol}           
VLT-spectropolarimetry of ESO 323-G077. The top panel gives the total 
flux spectrum 
$f_{\rm obs}$ 
(in $10^{-15}\,{\rm erg}\,{\rm s}^{-1}\,{\rm cm}^{-2}\,{\rm \AA}^{-1}$) 
of the bright core (solid line) and the estimated contribution of the 
host galaxy
$f_{\rm gal}$ (dashed line). The polarized flux 
spectrum $p_{\rm obs}\times f_{\rm obs}$ is plotted in the second panel.  
The relative linear polarization $p_{\rm obs}$ is displayed in the
third panel, and the 
polarization position angle $\theta$ in the bottom panel. The spectral
range $\lambda<5850$\AA\ is based on the G600B data while the G600R 
data are plotted for longer wavelength. The asterisks ($\ast$) 
at 4050~\AA\ mark an artifact due to the reflex in the G600B spectrum  
described in Sect.~2.} 
\end{figure*}

Figure~\ref{fig.pol}
combines the spectropolarimetric data of the G600B observation for
$\lambda<5850$~\AA\ and the G600R observation for $\lambda>5850$~\AA.
H$\alpha$ is truncated in order to show the weaker lines and
the continuum. The H$\alpha$ line profile structure is given in 
Fig.~\ref{fig.polline}(a). A flux spectrum without truncation
can be found in Fairall (\cite{fairall86}).  
The Tables~\ref{tab.line} and \ref{tab.cont} give flux values and
the polarization for emission lines and 250 -- 400~\AA\ wide
spectral intervals, respectively.

Due to the longer integration time the
signal to noise ratio of the G600B data is higher than for G600R. 
A small step in the percentage polarization $p$ and the polarized
flux $p\times f$ is visible at $\lambda=5850$~\AA . This is
probably due to slightly different seeing conditions and slightly different
slit positions for the two pointings, resulting in slightly
different relative flux contributions from
the polarized nucleus and the unpolarized host galaxy. The 
atmospheric dispersion corrector used with FORS helps to keep 
such effects small, but they cannot be completely eliminated. 
In our data, the polarization $p(\lambda)$ 
of the G600B observation is lower by $\Delta p=0.13~\%$, than for 
the G300V observation, while $p$ of G600R is 0.14~\% higher than for G300V. 
Therefore, whenever possible, mean values were calculated and given in 
Tables~\ref{tab.line} and \ref{tab.cont}. 

The wavelength region from 4050 to 4150~\AA\ in the G600B 
observation is affected by an optical reflex, which is known to occur 
for this
Grism/Wollaston configuration. The comparison with the 
reflex-free G300V observation indicates that only  
the flux spectrum and the relative polarization $p$ are affected, but not the 
polarized flux and the polarization angle.

In addition we observed spectropolarimetric standard stars 
(\object{HD 94851}, \object{HD 126593}, 
\object{BD+25$^\circ$727}\footnote{Note that in the list of 
polarized standard stars of 
Turnshek et al.~(\cite{turnshek90}) the polarization of 
BD+25$^\circ$727 should be $p=6.27~\%$ instead of 
$p=4.27~\%$ (e.g. Whittet et al.~\cite{whittet92}).}) 
for estimating the instrument polarization, which was found   
to be always 
$<$ 0.1~\% and thus negligible within the accuracy required for this study. 

%

\section{Observational results}
\subsection{The flux spectrum}
\label{sect.spec}
The observed combined flux and polarization spectra of  
ESO 323-G077 are displayed in Fig. 1.  
The  flux spectrum shows a flat or slightly reddish
continuum (in $F_\lambda$) and broad emission lines due to 
H\,{\sc i}, He\,{\sc i} and Fe\,{\sc ii}.
Very strong narrow lines are due to 
[O\,{\sc iii}] $\lambda\lambda$4959,5007. 
There are also strong emissions of
[O\,{\sc ii}] $\lambda$3726/29, 
[Ne\,{\sc iii}] $\lambda$3869, 
[S\,{\sc ii}] $\lambda\lambda$6716,6731, and there are traces of 
[N\,{\sc ii}] $\lambda$6583,6548, 
[O\,{\sc i}] $\lambda$6300, and
[O\,{\sc iii}] $\lambda$4363.
On top of the broad H\,{\sc i} line components weak, 
narrow H\,{\sc i} lines are clearly visible.

Measured fluxes  
(uncorrected for slit losses) for the narrow and broad lines 
and for 250--400\AA -wide flux intervals are
given in Tables \ref{tab.line}, and \ref{tab.cont}, respectively.
For the H\,{\sc i} lines it is difficult to separate accurately the
narrow and broad line components. Therefore, we present in 
Table~\ref{tab.line} the flux of the total H\,{\sc i} emission. 
However, the contribution of [N\,{\sc ii}] to H$\alpha$ 
and [O\,{\sc iii}] to H$\gamma$ has been subtracted. About 10~\%\ of the
total H$\alpha$ and  
H$\beta$ line flux are due to the narrow line components. This percentage is
uncertain by about a factor of 2 
due to  uncertainties in  the profiles adopted for 
the broad component, which has been inferred from the comparison of 
the line profiles in total and polarized flux 
(Fig.~\ref{fig.polline}, Sect.~\ref{sect.line}).

Our continuum and line fluxes for the stronger lines  
agree within about 20~\% with those reported by 
Winkler (\cite{winkler92b}). 


\begin{table} [t]
\caption{Observed emission line properties. The flux $f_{obs}$
is given in $10^{-15}\,{\rm erg}\,{\rm s}^{-1}\,{\rm cm}^{-2}$.
Upper $p_{\rm obs}$-limits are 
given for stronger lines without a significant polarization detection.
Note that the fluxes refer to a 1 arcsec slit with seeing losses.}
\label{tab.line}
\vspace*{-0.2cm}
\begin{center}       
\begin{tabular}{llcccc} 
\noalign{\hrule\smallskip}
$\lambda_{\rm obs}$ 
            & line  & $f_{\rm obs}$ & $p_{\rm obs}$ 
                           & $\theta_{\rm obs}$ & errors  \\
~[\AA]   &                 &  $[10^{-15}]$ & [\%] & [$^\circ$] 
                                     & $[\%/^\circ]$ \\ 
\noalign{\smallskip\hrule\smallskip}
3785.5   & [O\,{\sc ii}]   & 22   & $<3$  \\
3927.3   & [Ne\,{\sc iii}] & 16   & $<2$  \\
3948.9   & H8              & 8    \\
4027.5   & [Ne\,{\sc iii}] & 6    \\
4163.4   & H$\delta$       & 21   & 11.6  & 75.1  & 5.8/13.3 \\
4407.6   & H$\gamma$       & 79   & 9.8   & 81.4  & 1.9/5.5  \\
4427.9   & [O\,{\sc iii}]  & 4     \\
         & Fe\,{\sc ii}    & 298  & 6.7   & 81.7  & 1.0/4.2  \\
4926.1   & H$\beta$        & 389  & 6.4   & 83.2  & 0.5/2.2  \\
5034.5   & [O\,{\sc iii}]  & 60   & $<2.5$ \\
5083.5   & [O\,{\sc iii}]  & 213  & $<1.0$ \\
         & Fe\,{\sc ii}    & 265  & 5.6   & 84.8  & 1.1/5.6  \\
5965.5   & He\,{\sc i}     & 30$^a$ &     & 78.9  & $\phantom{0.0}$/8.1 \\
6398.6   & [O\,{\sc i}]    & 8     \\
6663.3   & H$\alpha$       & 2410 & 3.9   & 84.7  & 0.1/0.7  \\
6683.0   & [N\,{\sc ii}]   & 90   \\
6820.2   & [S\,{\sc ii}]   & 20   & $<1.5$ \\
6834.4   & [S\,{\sc ii}]   & 23   & $<1.5$ \\
\noalign{\smallskip\hrule\smallskip}
\end{tabular}
\end{center}
\end{table}

\begin{table} [h]
\caption{Measured mean flux $f_{\rm obs}$ 
(in $10^{-15}\,{\rm erg}\,{\rm s}^{-1}\,{\rm cm}^{-2}\,{\rm \AA}^{-1}$) 
and polarization $p_{\rm obs}$, $\theta_{\rm obs}$ of ESO 323-G077 for  
selected intervals in the observed wavelength frame. The flux and 
polarization of the active nucleus 
$f_{\rm agn}$, $p_{\rm agn}$ are derived after correction for the estimated 
dilution by the host galaxy and narrow lines. 
Note that the fluxes refer to a 1 arcsec slit with seeing losses.}
\label{tab.cont}
\vspace*{-0.2cm}
\begin{center}       
\begin{tabular}{lccccc} 
\noalign{\hrule\smallskip}
interval & $f_{\rm obs}$ & $p_{\rm obs}$ & $\theta_{\rm obs}$ &
                  $f_{\rm agn}$ & $p_{\rm agn}$   \\
$\lambda\lambda_{\rm obs}$  
         &        & [\%] & [$^\circ$] &           & [\%] \\
\noalign{\smallskip\hrule\smallskip}
3500--3750     &  4.2  &  7.77  & 83.5  &  3.3  & 9.89 \\
3750--4000     &  5.3  &  6.70  & 84.1  &  4.3  & 8.26 \\
4000--4250     &  6.0  &  5.85  & 84.7  &  4.6  & 7.63 \\
4250--4500     &  6.9  &  5.35  & 84.5  &  5.0  & 7.38 \\
4500--4850$^a$ &  7.7  &  4.70  & 84.0  &  5.4  & 6.70 \\
4850--5150$^b$ &  9.7  &  4.15  & 83.9  &  6.5  & 6.19 \\
5200--5600     &  8.4  &  3.94  & 84.1  &  5.9  & 5.60 \\
5600--6000     &  8.3  &  3.51  & 84.2  &  5.7  & 5.11 \\
6000--6400     &  8.7  &  3.30  & 84.7  &  6.0  & 4.79 \\
6400--6800$^c$ & 15.5  &  3.49  & 84.4  & 12.8  & 4.23 \\
6800--7200     &  8.8  &  2.74  & 85.0  &  6.1  & 3.95 \\
7200--7600$^d$ &  8.9  &  2.63  & 85.2  &  6.3  & 3.71 \\
7600--8000$^d$ &  8.6  &  2.46  & 85.1  &  6.1  & 3.47 \\
8000--8400$^d$ &  8.5  &  2.25  & 85.7  &  6.0  & 3.19 \\
\noalign{\smallskip\hrule\smallskip}
\end{tabular}
\end{center}
\vspace*{-0.6cm}
\begin{description}
\item[$a$:] includes the strong Fe\,{\sc ii} $\lambda$4550 emission  
\item[$b$:] includes the strong H$\beta$ and [O\,{\sc iii}] emission lines 
\item[$c$:] includes the strong H$\alpha$ line 
\item[$d$:] includes strong telluric absorption, which are not corrected for
     the $f_{\rm obs}$ and $f_{\rm agn}$ values
\end{description}
\end{table}

\subsection{Continuum and line polarization}

The solid curves in Fig.~\ref{fig.pol} show the spectroscopic and 
spectropolarimetric signal obtained for ESO 323-G077. The corresponding
polarization parameters, linear polarization $p_{\rm obs}$ and position 
angles $\theta_{\rm obs}$,  are listed in the Tables \ref{tab.line}
and \ref{tab.cont} for the emission lines and for 
$250-400$~\AA \ spectral intervals, respectively. 
The parameters $p_{\rm obs}$ and $\theta_{\rm obs}$ 
were calculated from the observed Stokes Q and U spectra according 
to $p = (Q^2+U^2)^{1/2}/f$, where $f$ is identical to Stokes $I$, 
and $\theta=0.5\cdot{\rm arctan}(U/Q)$. 

The errors of the continuum polarization (Table~\ref{tab.cont}) 
include the
instrumental polarization.  The errors of the 
polarization of the broad lines is strongly affected by the uncertainties 
in the continuum  definition. Therefore, the errors
are larger for weaker lines and for the broad Fe\,{\sc ii} features.
 
No significant polarization was measured for the narrow lines.
Therefore, the narrow lines are absent in the polarized 
flux spectrum $p\times f$. The polarization $p$ shows at the 
position of the narrow lines a sharp minimum due to the dilution
of the polarized continuum by the unpolarized narrow-line light. 
\smallskip

\noindent
The observed spectropolarimetric properties of ESO 323-G077 can be 
summarized as follows:

\begin{itemize}
\itemsep0cm
\item{} The observed linear polarization $p_{\rm obs}$ in the 
       continuum rises from 
       about 2.2~\% at 8300~\AA\ to 7.8~\% at 3600~\AA. 
       Compared to the total flux, the colours of the polarized 
       flux $p\times f$ are significantly bluer by about 
       ${\rm B-V} = 0.30$~mag. 
\item{} The broad H\,{\sc i} lines, but also the  
        Fe\,{\sc ii} features at $\lambda$5200 and $\lambda$4750
        and He\,{\sc i} $\lambda$5876, show a higher 
       polarization $p_{\rm obs}$ than the adjacent continuum.
       The polarization of the Balmer lines decreases from
       H$\delta$ over H$\gamma$, H$\beta$ to H$\alpha$.
\item{} The position angle of polarization is 
        $\theta=84^\circ\,(\pm 2^\circ)$, independent of wavelength.
\item{} The narrow [O\,{\sc iii}] lines and all other forbidden
        lines show no detectable polarization.
\end{itemize}

\subsection{The possible contribution of the Milky Way ISM to the
observed polarization and reddening}
\label{sect.ism}

Because of the relatively low Galactic latitude $b=+22.4^\circ$ 
of ESO 323-G077 an interstellar reddening and a possible Galactic 
interstellar contribution to the observed polarization 
has to be considered. According to the
far-IR dust emission 
(Schlegel et al.~\cite{schlegel98} / NED database) 
an interstellar reddening of $E_{\rm B-V}=0.10$ is to be expected
in the direction of ESO 323-G077.

Some information on the interstellar polarization in this direction
is provided by our 
observations of the star GSC 07777-00007 located only
30\arcsec\ from the nucleus of ESO 323-G077. For this star we find 
a continuum polarization of $p=0.22~\%$, $\theta=177^\circ$.
A literature search yields similar values for the  
interstellar reddening and polarization
of other distant stars in the general region of ESO 323-G077, e.g. for the
the two B stars \object{HD 112192} and 
\object{HD 114981} 
(Mathewson \& Ford \cite{mathewson70}; Hill \cite{hill70}; 
Kilkenny et al.~\cite{kilkenny75}). Hence we
conclude that the result for GSC 07777-00007 can be regarded as 
representative for this direction.

Our observed  spectrum indicates that
GSC 07777-00007 is a late F star, or an early metal deficient G star.
It seems safe to assume that such a star is intrinsically unpolarized. 
The observed spectrum and the apparent magnitude 
($m_V=11.4$~mag) yields a minimum distance $d>300$~pc and a minimum distance
to the galactic symmetry plane of about 150 pc. Hence the light path
to this star passes through most of the galactic dust layer. The observed
polarization of this star can, therefore, be regarded as
characteristic for the Galactic interstellar
polarization of extragalactic objects in this direction.
If this is the case, the Galactic
interstellar polarization along the line of sight is very small
compared to the intrinsic polarization of ESO 323-G077. 
Moreover, the position angle
of the interstellar polarization $\theta=177^\circ$ is practically 
perpendicular to the intrinsic polarization 
$\theta=84^\circ$. Hence, an interstellar polarization correction  
would enhance the intrinsic polarization by about 0.2~\% without
changing the angle significantly. Since systematic errors in the  
derivation of the relative contribution of the unpolarized light 
of the host galaxy to the polarized nuclear spectrum introduce larger
uncertainties, and since there may be local variations of the Galactic
interstellar polarization no attempt was made to correct for the 
small Galactic polarization.  

\subsection{Polarization structure of the emission lines}
\label{sect.line}

\begin{figure}
\begin{center}
\hskip-0.1cm\hbox{\psfig{figure=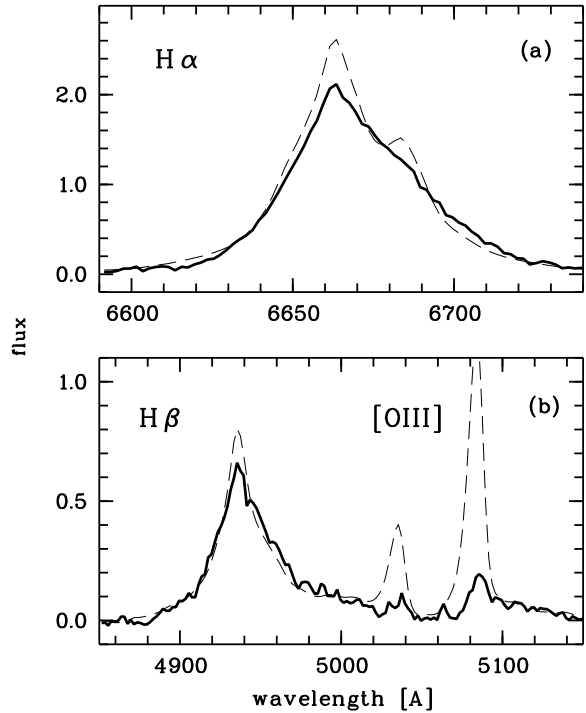,width=8cm}}
\end{center}
\vskip-0.5cm
\caption[] 
{\label{fig.polline}           
Line profiles of the {\bf (a)} H$\alpha$ and {\bf (b)} 
H$\beta$ and [O\,{\sc iii}] 
emission lines for polarized light $p\times f$ (solid line) 
and total light $f$ (dashed line). The flux scales 
(in $10^{-15}\,{\rm erg}\,{\rm s}^{-1}\,{\rm cm}^{-2}\,{\rm \AA}^{-1}$)  
refer to the $p\times f$ spectra. The $f$-spectra are scaled by
factors of 0.045 for H$\alpha$ and 0.07 for H$\beta$/[O\,{\sc iii}]
for this comparison of the broad line structure in $f$ and $p\times f$.} 
\end{figure}

Fig.~\ref{fig.polline} shows the line profiles for the total flux $f$ and
for the
polarized flux $p\times f$ of the strongest emission lines.
In each case the  continuum flux has
been subtracted. In the case of H$\alpha$ weak narrow nebular 
[N\,{\sc ii}] lines are superimposed. Apart from small effects
resulting from this blending, the line profiles of the
broad components of H$\alpha$ and H$\beta$ are obviously very
similar in the total and in the polarized flux. On the other hand, 
the narrow nebular lines [O\,{\sc iii}] 
near H$\beta$ and [N\,{\sc ii}] at 6683~\AA\ are very weak or absent
in the polarized light. The same seems to be true 
for the narrow line components of H$\alpha$
and H$\beta$. Thus, the line profile in polarized flux can be used 
as guide for separating the broad line components from the narrow 
line components in the total flux spectrum. We used this to estimate
the relative contribution of the narrow component to the total line
flux for H$\alpha$ and H$\beta$ (see Sect.~\ref{sect.spec}). 

Fig.~\ref{fig.polline} also indicates that the polarization in the
blue wing and the line center of the broad H$\alpha$ component
is about $p=4.5~\%$, while it is higher than 5~\% in the red
wing around 7000~\AA . For the broad H$\beta$ component the blue 
wing and line center have a polarization of about 7~\%, while the
red line wing shows about 8~\%. The polarization of the entire H$\alpha$
and H$\beta$ features as given in Table~\ref{tab.line} are lower
because of the contribution of the unpolarized narrow line components.  

Moreover we see for H$\alpha$ in the position angle spectrum in
Fig.~\ref{fig.pol} a small position angle rotation of 
about $8^\circ$ from the blue line wing to the red line wing. 
Such structures in the polarization angle are often seen 
in high quality data of the H$\alpha$-lines in Seyfert 1 galaxies
(e.g. Goodrich \& Miller \cite{goodrich94}; Young et al.~\cite{young99};
Schmid et al.~\cite{schmid00}; Smith et al.~\cite{smith02}).
It is beyond the scope of this paper to investigate the nature of this
effect. The interested reader is refered to the
detailed description and discussion in Smith et al.~(\cite{smith02}).

%

\section{Interpretation}


Seyfert 1 galaxies with high scattering polarization are rare 
(e.g. Smith et al.~\cite{smith02}). Good examples are Mrk 231, 
\object{Mrk 486}, \object{Mrk 704} and Fairall 51 
(Goodrich \& Miller \cite{goodrich94};
Smith et al.~\cite{smith95},\cite{smith97}; Schmid et al.~\cite{schmid01}).
ESO 323-G077 is a new member of this group. In the following
discussion we adopt for ESO 323-G077 the analysis procedure
developed and described by Schmid et al.~(\cite{schmid01})
for the high-polarization Seyfert 1 galaxy Fairall 51. This appears
reasonable since both objects have very similar spectropolarimetric
properties.   

Important for the spectropolarimetric analysis of Seyfert nuclei is
an estimate of the host galaxy's contribution to the intensity
spectrum. In contrast to F51 our spectra of ESO 323-G077 
show no undisturbed spectral features from
the host galaxy which could be used to estimate the
host galaxy's contribution. Therefore, we rely on the assumption 
that the broad lines and continuum from the nucleus have at a given
wavelength the same polarization, i.e. the relative polarization spectrum 
$p_{\rm AGN} = p_{\rm obs}\times f_{\rm obs}/f_{\rm AGN}$ is essentially
featureless apart from a slope from blue to red. This is expected if
the nuclear spectrum and the broad lines have the same scattering 
geometry, which is to be expected if the scattering occurs far away from the 
nuclear continuum and broad line emission region. 
On the basis of this assumption we subtracted from the observed spectrum 
a scaled ``typical'' galaxy spectrum, with the scaling factor adjusted 
to reach a polarization in the broad lines equal to the adjacent 
continuum. As a ``typical `` galaxy spectrum we used that of the 
E0 galaxy NGC 3379 (Kennicutt \cite{kennicutt92}), which should be 
a reasonable approximation for the central (bulge) region of
the ESO 323-G077 host galaxy. 
The relative flux of the host galaxy spectrum $f_{\rm gal}$ derived 
in this way has been plotted in the top panel of Fig. 1 as a dashed line.
Table~\ref{tab.cont} gives the resulting flux 
$f_{\rm AGN}=f_{\rm obs}-f_{\rm gal}-f_{\rm NLR}$ 
and polarization $p_{\rm AGN}$ for the active nucleus.
For this purpose we also subtracted the narrow line emission $f_{\rm NLR}$. 
   
According to the study on Fairall 51, the nuclear spectrum $f_{\rm AGN}$ 
can be further separated into direct $f_{\rm d}$ and scattered $f_{\rm s}$
light. These components must have the following properties in order 
to explain the presence of a high scattering polarization 
(see Schmid et al.~\cite{schmid01}):
\begin{itemize}
\itemsep0cm
\item{} $f_{\rm d}$ must pass through a dust layer producing 
        an attenuation and reddening of the direct (unpolarized) light,
\item{} a sufficiently large scattering region must be present and 
        visible from us which can produce the observed fraction 
        of polarized nuclear light,
\item{} the scattering angle has to be of the order $\approx 45^\circ$ 
        to produce a sufficiently high polarization.
\end{itemize}
  
The spectrum of  ESO 323-G077 obviously shows all spectral
components required by the model. The amount of the observed
polarization is slightly lower 
but of the same order as in the case of Fairall 51.
Hence, the relative contributions to the observed light are
similar. Thus, the new observations strongly support the
general scenario outlined for the case of Fairall 51 and illustrated by Fig.~6
of Schmid et al.(~\cite{schmid01}). In this model the circumnuclear
torus is inclined so that the direct light from the central source
has a line of sight close to the torus walls where substantial reddening
occurs. The high polarization is due to scattering by optically thin 
dust located in polar directions above and inside the torus. The scattering
angle and therefore the inclination of the torus plane must be around 
$\approx 45^\circ$ with respect to the line of sight in order to explain 
the high scattering polarization in ESO 323-G077 and Fairall 51. 
For smaller inclinations the scattering
angle would be too small to produce a strong polarization signal  
(as in normal Seyfert 1s), while for larger inclinations the
direct (unpolarized) emission from the continuum source and the BLR would be
completely hidden by the torus as in Seyfert 2s.

\begin{figure}
\begin{center}
\hskip-0.1cm\hbox{\psfig{figure=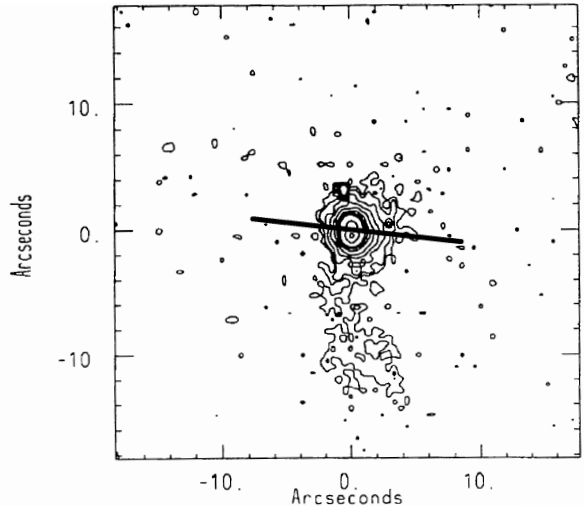,width=8cm}}
\end{center}
\vskip-0.5cm
\caption[] 
{\label{fig.oiii}           
Comparison of the orientation of the of the extended 
[O\,{\sc iii}] emission in ESO 323-G077 (map from 
Mulchaey et al.~\cite{mulchaey96a}) and of the orientation
of the observed scattering polarization.} 
\end{figure}

While the data described here support the qualitative 
model developed on the basis of Fairall 51, the new data of ESO323-G077 
contain important additional information concerning the scattering
geometry. Important is that the polarization position angle (PA) defines the 
relative orientation of the scattering region with respect to the 
light source, since the {\bf E} vector of the scattered light has   
to be perpendicular to the scattering plane (the 
plane containing the incident and scattered ray). As shown 
by Fig.~\ref{fig.oiii} the observed PA of 
polarization in ESO 323-G077 is within the limits of the observational
errors perpendicular to the PA of the extended [O\,{\sc iii}]
emission region observed by Mulchaey et al.~(\cite{mulchaey96a})
for this galaxy. This is exactly the orientation which 
we have to expect if the observed extended [O\,{\sc iii}] emission
originates in an ionization cone produced by the non-isotropic 
radiation field of the AGN. 

%

Such [O\,{\sc iii}] emission
regions are seen in many Seyfert 2 galaxies (e.g. Pogge~\cite{pogge89};
Mulchaey et al.~\cite{mulchaey96a},b) and are normally
taken as evidence for an emission of the ionizing radiation
from the active nucleus along the symmetry axis of the disk and  
torus system present according to the unified model of Seyfert galaxies 
(Antonucci \& Miller \cite{antonucci85};
Krolik \& Begelman \cite{krolik88}; Antonucci \cite{antonucci93}).  
Apart from Seyfert 2 galaxies the perpendicular alignment between 
the polarization and the objects symmetry axis has also been 
observed in narrow-line
radio galaxies (the AGN type 2 equivalent to the type 1 broad-line radio
galaxies, see e.g. Cohen et al.~\cite{cohen99}).  

In Seyfert 1 galaxies the situation is more complicated.
For those low polarization Seyfert 1s where radio structures or
extended emission line regions have been detected the position
angle of polarization can be parallel or perpendicular to these structures
(Smith et al.~\cite{smith02}). 

What is the situation for Seyfert 1s with high scattering polarization?
In Mrk 231 the polarization is clearly perpendicular to a well
defined radio structure (Goodrich \& Miller \cite{goodrich94}; 
Neff \& Ulvestad \cite{neff88}). With ESO 323-G077 we have now another 
case of a high polarization Seyfert 1 galaxy which shows a 
perpendicular alignment between polarization and extended nebular emission. 
We are not aware that extended radio structures or emission line regions 
have been detected in other systems of this class. 
Thus, the high polarization
Seyfert 1 galaxies ESO 323-G077 and Mrk 231 behave like Seyfert 2 galaxies.
Therefore, they strongly support the dust torus induced ionization cone model 
within the unification scheme which suggest that Seyfert 1 and 2 
nuclei are the same physical objects viewed at different orientations. 

Hence, the new data and in particular the  
orientation of the polarisation with respect to the observed [O\,{\sc iii}]
emission strongly support our assumptions that the high-polarization Seyfert 1 
galaxies, like ESO 323-G077, Fairall 51, Mrk 231 and probably others, 
are AGN with orientations very close to the transition between 
Seyfert~1 and Seyfert~2. 
In these borderline Seyfert 1 objects a major fraction 
of the observed light from the nucleus is scattered light reaching us 
via illuminated dust regions. These could be optically thin dust clouds
within or above the torus and/or the illuminated inner edge of the
dust torus itself. The other part of the light 
reaches us directly but significantly reddened by dust extinction in
the semi-transparent inner boundary region of the torus.   

\section{Conclusions}

We have shown that the high polarization of the Seyfert 1 galaxy 
ESO323-G077 has very similar properties as observed earlier in
Fairall 51. As for Fairall 51, the observations of ESO323-G077 can 
be understood assuming that the AGN is observed at an inclination of 
roughly $\approx 45^\circ$, so that the nucleus is partially obscured
by the torus and where scatterings by dust in polar directions
above or within the torus produce the high linear polarization.

The fact that in the case of ESO 323-G077 the observed polarization 
is found to be perpendicular to the AGN's projected symmetry axis,
as deduced from the orientation of the extended [O\,{\sc iii}]
emission region, supports the dust torus induced ionization
cone model. 

The finding that the inclination
angle of the active nuclei in ESO 323-G077 and Fairall 51 
cannot be much different from 45 degrees, 
provides important constraints on the aspect ratio of the AGN dust
tori. Further information on this question could be derived from
the frequency of objects like ESO323-G077, Fairall 51, and Mrk 231, 
which must be regarded in the unified model for Seyfert galaxies as 
borderline cases between type 1 and type 2 systems. 
Because of the constraints on the inclination angles which can be derived 
from the amount of observed polarization, these high polarization
Seyfert 1 galaxies can play an important role for determining the
geometric properties of AGN. Therefore, it will be very desirable to
look for further examples of this apparently rare class of Seyferts
in order to obtain a meaningful sample for a statistical
evaluation of their properties. 

\begin{acknowledgements}
We thank the referee, A. Robinson, for thoughtfull comments and for
pointing out the observations of the radio structure in Mrk 231 
in the literature.  

\end{acknowledgements}

\end{document}